\documentclass[openacc]{rstransa}

\usepackage{graphicx}
\usepackage{bm}
\usepackage{amsmath}
\usepackage{dsfont}

\begin{document}

\title{Quasiperiodic granular chains and Hofstadter butterflies}

\author{
Alejandro J. Mart\'inez$^{1}$, Mason A. Porter$^{2}$, and P. G. Kevrekidis$^{3}$}

\address{$^{1}$Oxford Centre for Industrial and Applied Mathematics, Mathematical Institute, University of Oxford,
Oxford OX2 6GG, UK\\
$^{2}$Department of Mathematics, University of California, Los Angeles, CA 90095, USA\\
$^{3}$Department of Mathematics and Statistics, University of Massachusetts, Amherst, MA, 01003, USA}

\subject{Nonlinear waves, nonlinear dynamics, dynamical systems, condensed-matter physics}


\corres{Mason A. Porter\\
\email{mason@math.ucla.edu}}

\begin{abstract}
  We study quasiperiodicity-induced localization of waves in strongly precompressed granular chains. We propose three different setups, inspired by the Aubry--Andr\'e (AA) model, of quasiperiodic chains; and we use these models to compare the effects of on-site and off-site quasiperiodicity in nonlinear lattices. When there is purely on-site quasiperiodicity, which we implement in two different ways, we show for a chain of spherical particles that there is a localization transition (as in the original AA model). However, we observe no localization transition in a chain of cylindrical particles in which we incorporate quasiperiodicity in the distribution of contact angles between adjacent cylinders by making the angle periodicity incommensurate with that of the chain. For each of our three models, we compute the Hofstadter 
  spectrum and the associated Minkowski--Bouligand fractal dimension, and we demonstrate that the fractal dimension decreases as one approaches the localization transition (when it exists). Finally, in a suite of numerical computations, we demonstrate both localization and that there exist regimes of ballistic, superdiffusive, diffusive, and subdiffusive transport. Our models provide a flexible set of systems to study quasiperiodicity-induced analogs of Anderson phenomena in granular chains that one can tune controllably from weakly to strongly nonlinear regimes.
\end{abstract}



\maketitle


\section{Introduction} \label{sec1}

Quasicrystals are solids whose structure is ordered but not periodic~\cite{Janot}. For many years, it was thought that quasicrystals could only be produced artificially. However, almost a decade ago, the first natural quasicrystal was found in Russia~\cite{NaturalQuasi}. A common type of quasicrystal arises when atoms are arranged so that they possess symmetries, such as $5$-fold symmetry, that are 
forbidden to periodic crystals\footnote{By the \textit{crystallographic restriction}, crystals can only have certain rotational symmetries: $2$-fold, $3$-fold, $4$-fold, or $6$-fold symmetry~\cite{Definition}.}. A famous two-dimensional (2D) example is a Penrose tiling~\cite{Penrose}. More generally, quasiperiodic structures can exist in any number of dimensions as structures 
with a broken translational symmetry. In 1D, the most common models used in the study of quasiperiodic systems are
a Fibonacci quasicrystal~\cite{Fibo} and the Aubry--Andr\'e (AA) model~\cite{AAmodel}. These models are topologically equivalent 
to each other~\cite{Equivalence}, in the sense that it is possible to continuously deform one into another without closing any bulk gaps.

A key feature of quasiperiodic potentials, arising prominently in the study of Schr{\"o}dinger equations~\cite{AAmodel}, is the transition from a ``metallic'' phase, in which all eigenstates are delocalized, to an ``insulating'' phase, in which they are localized. See, e.g., the analysis in~\cite{fishman,aullbach} and the experiments in~\cite{OAA}. It is of considerable interest to extend these studies in various ways, including to nonlinear systems (see, e.g., the work of~\cite{flach} and references therein), many-body systems, discrete systems, and settings with controllable interactions. For example, there have been relevant investigations in both bosonic and fermionic settings~\cite{refael,mastropietro}.

In the present paper, we study the effects of quasiperiodicity in strongly precompressed granular chains~\cite{Hertz1,Hertz2}, in which neighboring particles interact with each other according to a Hertzian potential.
Our aim is to illustrate that localization of eigenmodes can occur in quasiperiodic granular chains and to explore the conditions --- with respect to both models and experimental setups --- in which it occurs. We illuminate these conditions by comparing a variety of different models with one or both of on-site and off-site quasiperiodic structures. As demonstrated in a wealth of research over more than three decades \cite{Hertz3}, granular chains are extremely versatile, as one can adjust interaction potentials; readily tune them between almost linear, weakly nonlinear, and strongly nonlinear regimes by applying different precompression strengths; construct them using particles of different sizes, shapes, and material properties; and so on. This has yielded a wealth of insights about a diverse set of physical phenomena, including acoustic realizations of many concepts from condensed-matter physics \cite{Hertz4}. Most centrally for our discussion, this includes the dynamics of wave transport and localization in disordered nonlinear systems in both theoretical ~\cite{DisorderTheo1,DisorderTheo2} and experimental~\cite{DisorderExp} studies. Other phenomena (both theoretical and applied) from condensed-matter and quantum physics that have been realized in granular crystals include an analog of a Ramsauer--Townsend resonance in the form of a square well \cite{ajm2016b}, switching and rectification \cite{Nature11}, and others. More broadly, granular chains provide a wonderful playground that enables systematic exploration of the role of lattice structure (e.g., material heterogeneity~\cite{vakbook,katjarev}) and fundamental dynamic (e.g., rogue waves~\cite{sen2}) and thermodynamic (e.g., equipartition~\cite{sen3}) phenomena.

The remainder of our paper is organized as follows. In Section \ref{sec2}, we briefly review the AA model. In Section \ref{sec3}, we present three models of 1D quasiperiodic lattices: two with on-site quasiperiodicity and one with off-site quasiperiodicity. In Section \ref{sec4}, we linearize the governing equations of our three models. In Section \ref{sec5}, we demonstrate that such models may possess a Hofstadter butterfly structure. In Section \ref{sec6}, we examine energy transport and localization in our models. We conclude and suggest some interesting directions for future research in Section \ref{sec7}.


\section{A brief review of the Aubry--Andr\'e model} \label{sec2}

The prototypical form of the Aubry--Andr{\'e} (AA) model at the level of a tight-binding model is
\begin{equation} \label{aa-orig}
	E \Psi_n= \Psi_{n+1}+\Psi_{n-1}+\lambda V(nq+r)\Psi_n \,,
\end{equation}
where $\Psi$ is a wavefunction, $n$ indexes the lattice site, $E$ is energy, and $V$ is a potential. We suppose that the on-site energy is modulated by a lattice distortion with wave vector $q\in\mathds{R}$, which is incommensurate with $2\pi$ and has phase $r$. We also suppose that $V(x)=V(x+2\pi n)$, where $x\in \mathds{R}$ and $n\in \mathds{Z}$. 

In \cite{AAmodel}, Aubry and Andr\'e proved several fundamental properties of the eigenmodes of \eqref{aa-orig}. They showed that a ground state exists and that it undergoes a transition from analyticity for $\lambda<\lambda_c$ to nonanalyticity for $\lambda>\lambda_c$ for certain $\lambda_c(q)$ and when $q$ is not a Liouville number. That is, $q\in \mathds{R}\setminus \mathds{Q}$ and there exist $\gamma$ and $r$ such that
\begin{equation}
	 \left|\frac{q}{2\pi} - \frac{p_1}{p_2}\right|> \gamma\frac{1}{p_2^r}
\end{equation}
is satisfied for any rational number $p_1/p_2$.

Aubry and Andr\'e also showed, using a perturbative approach, that the analyticity-breaking causes the eigenmodes of \eqref{aa-orig} to have very rich spatial properties. Specifically, one can write the eigenmodes of the modulated system as
\begin{equation}
	 \Psi_n (k) = e^{ikn} + \lambda \sum_{m=-\infty}^{\infty} \frac{v_m e^{im(qn + h)}}{2[\cos(mq+k)-\cos(k)]}\,,
\label{expansion}
 \end{equation}
where $v_m$ are the coefficients of the Fourier expansion of $V$. The eigenmodes and eigenvalues for $\lambda = 0$ are given by $e^{ikn}$ and $2\cos(k)$, respectively.
For the series in \eqref{expansion} to be convergent, one needs to satisfy a Diophantine condition. That is, there exist two positive constants $K$ and $\beta$ such that 
\begin{equation}
	 \left|\frac{k}{\pi} + m\frac{q}{2\pi} - n\right| > \frac{K}{m^{1+\beta}}
\end{equation}
for any integers $m$ and $n$. 

For the sake of simplicity, consider the special case in which the phase $r=0$ and $V(\xi n)=\cos(2\pi \xi x)$, where $\xi$ equals the golden ratio $(1+\sqrt{5})/2$. So, Eq.~\eqref{aa-orig} takes the form: $E \Psi_n= \Psi_{n+1}+\Psi_{n-1}+\lambda \cos(2\pi\xi n)\Psi_n$. In this case, Aubry and Andr\'e showed that for $\lambda >\lambda_c = 2$, all of the eigenmodes of \eqref{aa-orig} are exponentially localized, as in the Anderson model (in which the potential $V$ is disordered rather than quasiperiodic)\cite{anderson1958}, with the same characteristic localization length
\begin{equation}
	 \zeta = \frac{1}{\ln \left(\frac{\lambda}{2}\right)}\,.
\end{equation}
That is, $\Psi_n$ decays asymptotically as $e^{-n/\zeta}$ as $n \rightarrow \infty$. However, when $\lambda < 2$, most eigenmodes are given by extended, modulated plane waves. Interestingly, this implies that the loss of analyticity
is also associated with a transition at $\lambda_c$ to spatial localization of the eigenmodes, where $\lambda_c = 2$ in this specific example. This transition is called a {\it localization transition} or {\it Aubry--Andr\'e transition}. This result is a generic phenomenon in Schr\"odinger lattices, and it is thus relevant for a diverse variety of physical systems~\cite{OAA,Martinez:PRA2012,AABEC,AAPT}, including photonic lattices, Bose--Einstein condensates, and many others. Additionally, the spectra of the corresponding
Schr{\"o}dinger operators (with two or even more frequencies) is a topic of intense mathematical interest; see,
e.g.,~\cite{schlag} and references therein.


\section{Implementing the AA model in granular chains}  \label{sec3}

Several recent studies have generalized conventional granular chains in various ways. They have yielded several interesting insights, and they promise to result in a host of others in the coming years \cite{Hertz3,Hertz4}. One type of a generalized granular chain is a {\it cradle system}~\cite{cradle}, in which particles are attached to linear oscillators, enabling the use of on-site potentials in a way that is independent of particle--particle interactions. Several potential realizations of such a setting were given
in~\cite{cuevas} (although they have yet to be implemented experimentally, to the best of our knowledge).
Another fascinating variant arises from examining a chain of particles with internal resonators, such as by coupling a secondary particle inside a principal one. This leads to a locally-resonant granular chain, which is sometimes called a {\it mass-in-mass} (or {\it mass-with-mass}, if the secondary particle is external) 
chain~\cite{mim,mim1,mim2}. Additionally, the use of particles with non-spherical geometries can drastically modify particle--particle interactions. For instance, with cylindrical particles, 
although one has the same functional relation between 
the force and the displacements as with spherical particles, one can tune the magnitude of such interactions by changing the contact angle between adjacent cylinders~\cite{cili1,fli}. 

We will consider granular chains with all of the above types of variations. The equations of motion in our general setting are
\begin{align}
	\ddot{u}_n &= \underbrace{\alpha_n[\delta_n+u_{n-1}-u_n]_+^{3/2}-
	\alpha_{n+1}[\delta_{n+1}+u_n-u_{n+1}]_+^{3/2}}_\text{Hertzian interaction}\nonumber\\
	&\qquad -\underbrace{\beta_n u_n}_\text{elastic restitution} -\underbrace{ \gamma_n (u_n - v_n)}_\text{mass-in-mass interaction}\,,\label{3.1}\\
	\ddot{v}_n &= \gamma_n(u_n - v_n)\,,\label{3.2}
\end{align}
where $u_n$ is the displacement of the $n$th particle (where $n\in \{1,2,{\ldots},N\}$) measured from its equilibrium position in the initially compressed chain, $v_n$ is the displacement of the $n$th interior mass (when one particle is located inside another), and 
\begin{equation}
	\delta_n = \left(\frac{F_0}{A_n}\right)^{2/3}
\end{equation}
is a static displacement for each particle that arises from the static load $F_0 =
\text{const}$. 
There is a Hertzian interaction between a pair of particles only when they are in contact, so each particle is affected directly only by its nearest neighbors and experiences a force from a neighbor only when it overlaps with it. This yields 
\begin{equation}
[x]_+ = \left\{
	\begin{array}{lcc}
		x\,, & \text{if} & x>0\\
		0\,, & \text{if} & x\leq0
	\end{array}\right.\,.
\end{equation}

In our subsequent discussions (see Section~\ref{sec3}), we will consider various special cases of \eqref{3.1} and \eqref{3.2}, depending on the type of particle that we use to construct chains. Specifically, we work with three different models: two of them have spherical particles, and one has cylindrical particles. For our analysis and computations, we assume that $\alpha_n$, $\beta_n$, and $\gamma_n$ vary sinusoidally in space according to the following formulas:
\begin{align}
	 \alpha_n &= \bar{\alpha}_1+\bar{\alpha}_2\cos(2\pi n \xi)\,, \label{3.6} \\
         \beta_n &= \bar{\beta}\left[1+\cos(2\pi n \xi)\right]\,,\label{3.7} \\
	 \gamma_n &= \bar{\gamma}\left[1+\cos(2\pi n \xi)\right]\,, \label{3.8}
\end{align}
where $\xi$ is the golden mean $\frac{\sqrt{5}+1}{2}$ (unless we explicitly state otherwise), $\bar{\alpha}_1>\bar{\alpha}_2\geq 0$, and $\bar{\beta}$, $\bar{\gamma}\geq 0$. 
For simplicity, we separately examine the effects of \eqref{3.6}--\eqref{3.8}. This leads to three different models in which it may be possible to observe an AA transition. In Eqs.~\eqref{3.6}--\eqref{3.8}, $\bar{\alpha}_2$, $\bar{\beta}$, and $\bar{\gamma}$ are the quasiperiodic parameters that determine the strengths of the modulations for the different terms in Eqs.~\eqref{3.1} and~\eqref{3.2}.


\subsection{On-site quasiperiodicity: Two different variants of the AA model using spherical particles}\label{onsite}

In this section, we discuss the effect of an on-site quasiperiodicity on the dynamics of a granular chain by considering chains of spherical particles with local potentials. We set $\bar{\alpha}_2=0$ in Eq.~\eqref{3.6}, so the coupling parameter $\alpha_n$ is given by $\alpha_n = A_n/m_n$, where
\begin{equation}
	A_n =
	\frac{4E_{n-1}E_{n}\left(\frac{R_{n-1}R_n}{R_{n-1}+R_n}
	\right)^{1/2}}{3\left[E_{n}(1-\nu_{n-1}^2)+E_{n-1}(1-\nu_{n}^2)\right]}\,,\label{An}
\end{equation}
and the $n$th particle has elastic modulus $E_n$, Poisson ratio $\nu_n$, radius $R_n$, and mass $m_n$. We assume that the particles are identical, so $E_n=E$, $\nu_n = \nu$, $R_n=R$, and $m_n=m$. This, in turn, implies that $\alpha_n = \bar{\alpha}_1$, and we let $\bar{\alpha}_1=1$ without loss of generality.

\begin{figure*}
\centering
\includegraphics[height=8cm]{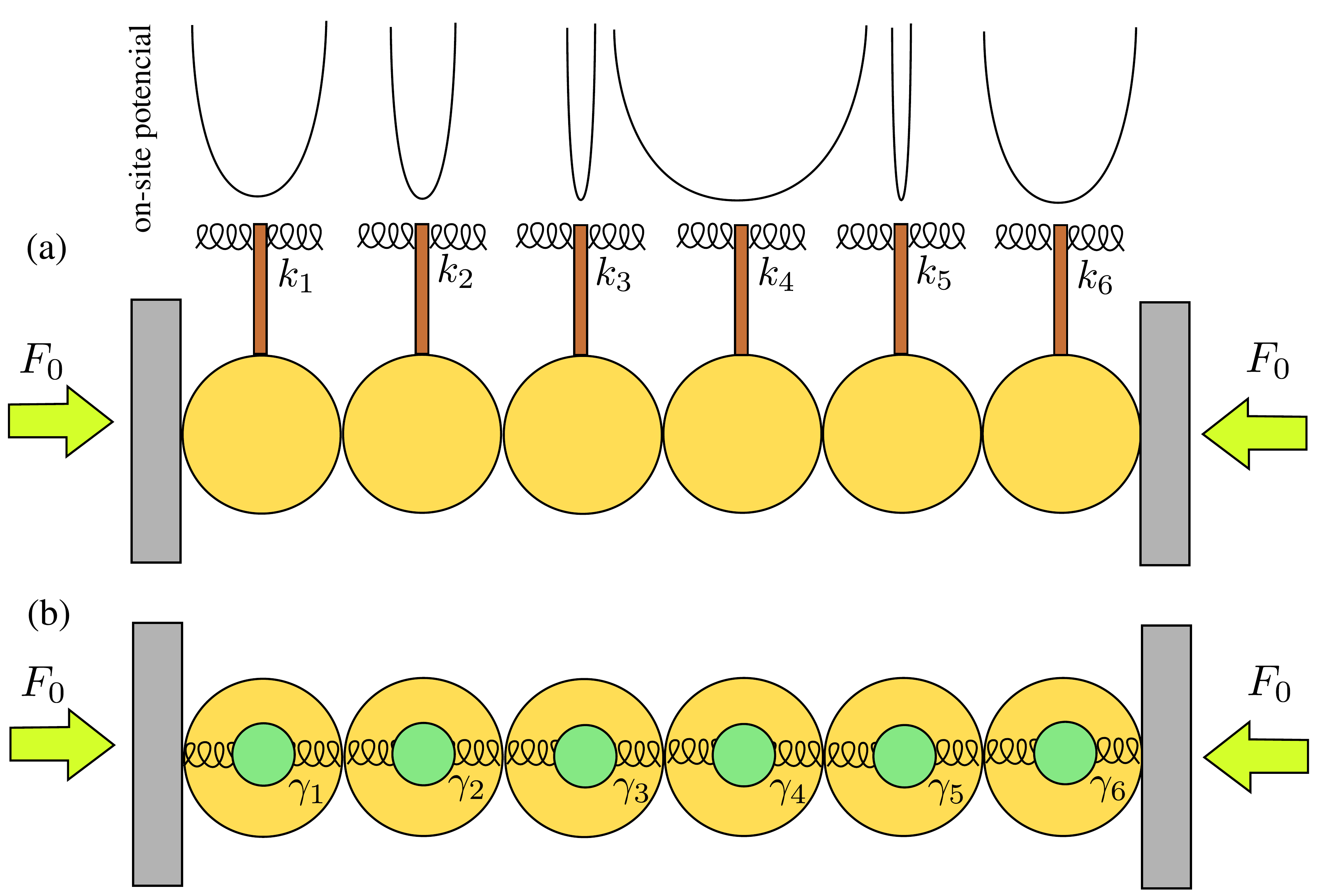}
\caption{Schematics of (a) model Ia and (b) model Ib.}
\label{fig1}
\end{figure*}


\subsubsection{Model Ia: $\bar{\beta}\neq 0$ and $\bar{\alpha}_2=\bar{\gamma} = 0$}\label{mod1}

Suppose that the particles in the chain are attached to a mechanical restitution mechanism, such that there is a linear force in the equations of motion (\ref{3.1},\ref{3.2}). In the limit of small angles, this can describe the well-known Newton's cradle, a system in which particles are aligned in one dimension and are suspended from a ceiling by strings so that the particles collide with each other along one dimension and oscillate. In the top panel of Fig.~\ref{fig1}, we show a schematic of this system. Studies of this system have focused primarily on waves that arise by releasing one of the particles at one end with some velocity~\cite{cuevas}. This produces a transfer of energy across the chain in the form of traveling waves~\cite{cradle}. Mulansky and Pikovsky studied disorder in closely related (nonlinearly coupled, and locally linear or nonlinear) oscillator systems~\cite{Mulansky:NJP2013}, showing numerically and using a fractional-nonlinear-diffusion approach that energy transport is subdiffusive. This helps further motivate the study of modulated systems, such as the AA model, in granular chains. In our case, the equation of motions are 
\begin{align}
	\ddot{u}_n &= [\delta+u_{n-1}-u_n]_+^{3/2}-[\delta+u_n-u_{n+1}]_+^{3/2}-\bar{\beta}\left[1+\cos(2\pi n \xi)\right] u_n\,,
\label{AA2}
\end{align}
where $\bar{\beta}>0$. In other words, the Hookean spring constants are positive for all $n$.


\subsubsection{Model Ib: $\bar{\gamma}\neq 0$ and $\bar{\alpha}_2=\bar{\beta} = 0$}\label{mod2}

We now consider chains composed of particles that include an internal degree of freedom (i.e., mass-in-mass particles)~\cite{mim}. Previous studies have focused on the generation~\cite{mim1,mim2}, of such lattices, their traveling-wave solutions~\cite{haitao,physd},
and their (bright and dark) breather-like excitations~\cite{anna1,anna2}. These works illustrate that coherent structures and their dynamics are enriched significantly by the presence of the internal degree of freedom (DOF). To give just one example, incorporating an internal DOF in the particles can lead to nonlocal solitary waves with non-vanishing tails (so-called ``nanoptera''), which have been observed experimentally in woodpile granular chains~\cite{jinkyu}. In our setting, we envision embedding a particle in the interior of each host particle, such that a particle and its interior mass are coupled via a linear restitution mechanism (such as a Hookean spring). 

The equations of motion, upon quasiperiodic modulation of the mass-in-mass
(MiM) resonator, are
\begin{align}
	\ddot{u}_n &= [\delta+u_{n-1}-u_n]_+^{3/2}-[\delta+u_n-u_{n+1}]_+^{3/2} -\bar{\gamma}\left[1+\cos(2\pi n \xi)\right] (u_n - v_n)\,,\\
	\ddot{v}_n &= \bar{\gamma}\left[1+\cos(2\pi n \xi)\right](u_n - v_n)\,,
\label{AA3}
\end{align}
where $\bar{\gamma}>0$. 


\subsection{Off-site quasiperiodicity: The AA model with cylindrical particles}\label{offsite}

Another approach for incorporating quasiperiodicity in a granular chain is by tuning particle--particle interactions. 
In existing experimental setups, to implement an AA model, one can use chains of cylindrical particles (rather than spherical ones). 
We are motivated by recent experiments~\cite{cili1}, where it was demonstrated that cylindrical particles
offer more flexibility than spherical particles for tuning particle--particle interactions, as one can change the contact angle between contiguous cylinders. Moreover, spatial and even temporal (periodic) variation of
contacts between cylindrical particles has been
proposed as an efficient way for implementing various functionalities,
including that of an acoustic transistor in~\cite{fli}.

\begin{figure*}
\centering
\includegraphics[height=8cm]{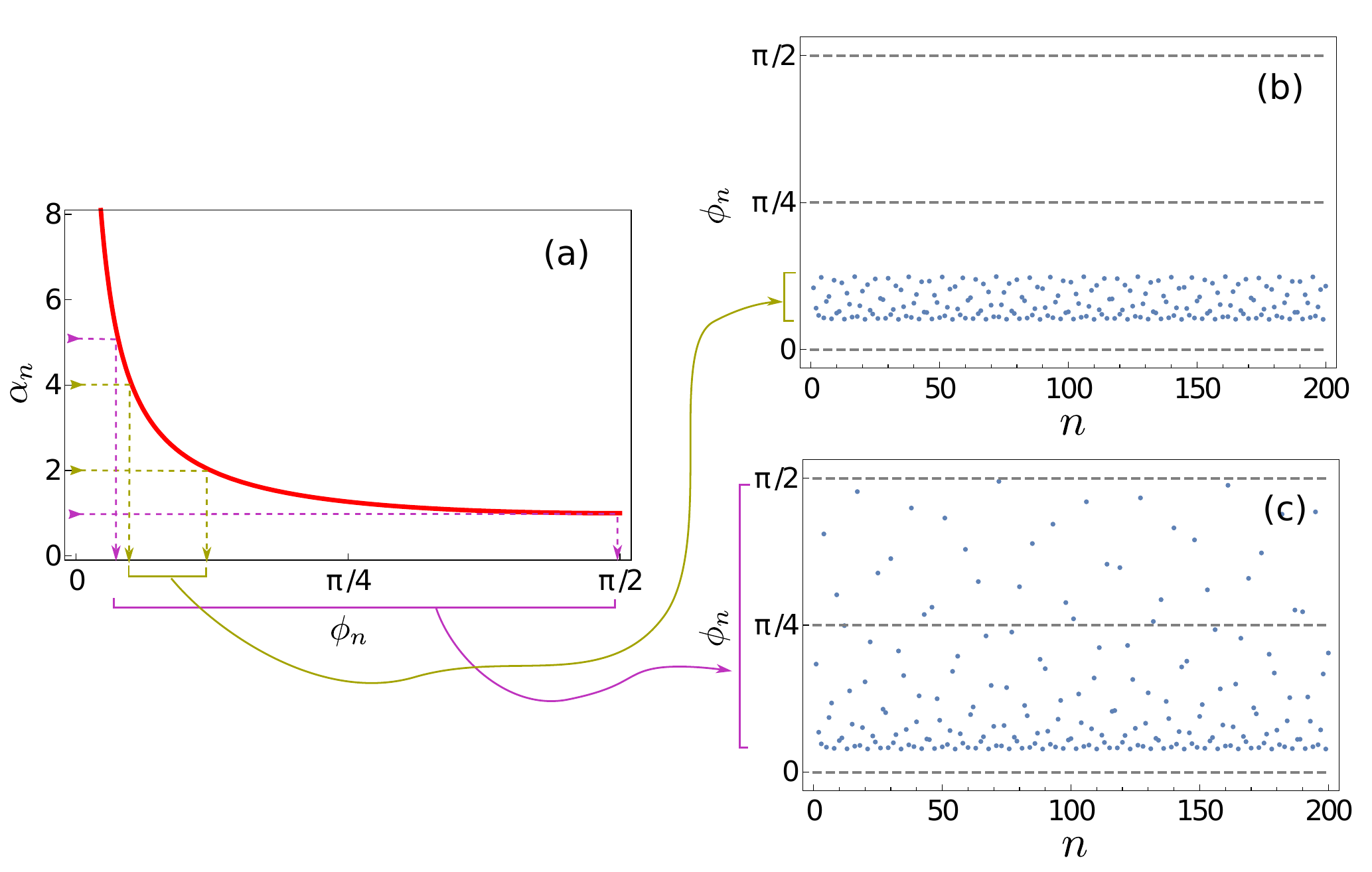}
\caption{(a) Interaction coefficient $\alpha_n$ for model II as a function of the contact angle $\phi_n$ between adjacent cylinders. In panels (b) and (c), we show the contact-angle distributions for two cases: (i) $\bar{\alpha}_1 = 3$ and $\bar{\alpha}_2 = 1$
and (ii) $\bar{\alpha}_1=3$ and $\bar{\alpha}_2 = 3$. We use arrows to represent the mapping process that we describe in the text.}
\label{Figcili1}
\end{figure*}


\subsubsection{Model II: $\bar{\alpha}_2\neq 0$ and $\bar{\beta} = \bar{\gamma} =0$}\label{mod3}

To give equations of motion for a chain of cylindrical particles, we first need to know the form of the Hertzian coefficient in this case. For identical cylinders, the interaction coefficients are~\cite{Johnson:Book}
\begin{equation}
	 \alpha_n(\phi_n) = \frac{Y}{m} g_1(\phi_n)\left[g_2(\phi_n)\sqrt{g_3(\phi_n)}\right]^{1/2}\,,
\label{ciliforce}
 \end{equation}
where $Y$ depends on the physical parameters of the particles, $m$ is the mass of a particle, and $\phi_n$ is the contact angle between cylinders $n-1$ and $n$ and it is defined $\text{mod}\,\pi/2$. Explicit forms of $Y$ and $g_i$ are
\begin{align*}
	 Y &= \frac{2E \sqrt{R}}{3(1-\nu^2)}\,,\\
 g_1(\phi_n) &= \sqrt{\frac{1}{\sin(\phi_n)}}\left(\frac{2K\left[e^2(\phi_n)\right]}{\pi}\right)^{-3/2}\,,\\
 g_2(\phi_n) &= \frac{4}{\pi e^2(\phi)}\,,\\
	 g_3(\phi_n) &= \left(\frac{a^2}{b^2}E\left[e^2(\phi_n)\right] - K\left[e^2(\phi_n)\right]\right)\left(K\left[e^2(\phi_n)\right] -E\left[e^2(\phi_n)\right]\right)\,,
\end{align*}
where $E$ is the elastic modulus, $\nu$ is the Poisson ratio, and $R$ is the radius of the circular cross section of the cylinders. The functions $K$ and $E$, respectively, are the complete elliptical integrals of 
the first and second kinds \cite{DLMF}. They are 
\begin{align*}
	 K(k) &= \int_0^{\pi/2} \frac{d\theta}{\sqrt{1-k^2 \sin^2\theta}}\,,\\
	 E(k) &= \int_0^{\pi/2} \sqrt{1-k^2 \sin^2\theta} d\theta\,,
\end{align*}
where $e = \sqrt{1-(b/a)^2}$ is the eccentricity of the contact area, $a$ is the semi-major axis length of the ellipse, and $b$ is its semi-minor axis length. One can approximate the quotient $b/a$ by $\left[(1-\cos\phi_n)/(1+\cos\phi_n)\right]^{2/3}$.

This yields the following equations of motion:
\begin{align} \label{aa3}
	\ddot{u}_n = \alpha_n(\phi_n)[\delta_n+u_{n-1}-u_n]_+^{3/2}-
	\alpha_{n+1}(\phi_n)[\delta_{n+1}+u_n-u_{n+1}]_+^{3/2}\,.
\end{align}
One can then control the interactions between particles by changing $\phi_n$~\cite{fli}.
This raises the following question of what the distribution of contact angles $\{\phi_n\}$ has to be to obtain $\alpha_n(\phi_n)= \bar{\alpha}_1 + \bar{\alpha}_2\cos(2\pi \xi n)$.
We address this issue by numerically inverting Eq.~\eqref{ciliforce}, so that quasiperiodic variation of $\alpha_n$ yields a quasiperiodic variation of angles in the interval $(\phi_{\text{min}},\phi_{\text{max}})$. In Fig.~\ref{Figcili1}, we show the contact-angle distributions generated in two cases: (i) when $\bar{\alpha}_1 = 3$ and $\bar{\alpha}_2 = 1$ and (ii) $\bar{\alpha}_1 = 3$ and $\bar{\alpha}_2 = 3$. 
We also note that $\alpha(\phi_n)\rightarrow \infty$ as $\phi_n\rightarrow 0$ and that $\alpha_n(\phi_n)$ has a lower bound at $\phi_n = \pi/2$. 


\section{Linear approximation} \label{sec4}

Depending on the strength of the precompression that we apply to a granular chain and the magnitude of the strains that arise in (or are exerted on) the chain, one can expand the Hertzian force in a power series about the equilibrium state. This process reduces the equations of motion for the granular chain to ones that resemble those for a Fermi--Pasta--Ulam--Tsingou (FPUT) chain \cite{Hertz4}. In particular, if the precompression is strong enough --- specifically,
if $\delta_n\gg |u_{n-1}-u_n|$ for all $n$ --- the dominant terms are the linear ones, so we linearize Eq.~\eqref{aa3} to get
\begin{align}
	\ddot{u}_n &= B_{n}u_{n-1}+B_{n+1}u_{n+1}-(B_n+B_{n+1}+ \beta_n+\gamma_n)u_n+\gamma_n v_n\,,\label{Linear}\\
	\ddot{v}_n &= \gamma_n(u_n-v_n)\,,\nonumber
\end{align}
where
\begin{equation}\label{equ9}
	B_{n} =\frac{3}{2}\alpha_n\delta_n^{1/2}\,. 
\end{equation}
By considering (without loss of generality at the level of this linear approximation) a complex representation of the wavefunctions, $u_n = \phi_n e^{i \omega t}$ and $v_n = \psi_n e^{i \omega t}$, and defining $E = -\omega^2$, we obtain
\begin{align}
	E \phi_n &= B_{n}\phi_{n-1}+B_{n+1}\phi_{n+1} -(B_n+B_{n+1}+ \beta_n+\gamma_n)\phi_n+\gamma_n \psi_n\,,\label{Eigenvalues}\\
	E \psi_n &= \gamma_n(\phi_n-\psi_n)\,,\nonumber
\end{align}
an eigenvalue problem that we can solve numerically by diagonalization. Using this linear description, we can now address the issue
of a localization (``metal--insulator'') transition for suitable incommensurate periodic coefficient variations of different types.

\begin{figure*}
\centering
\includegraphics[height=8cm]{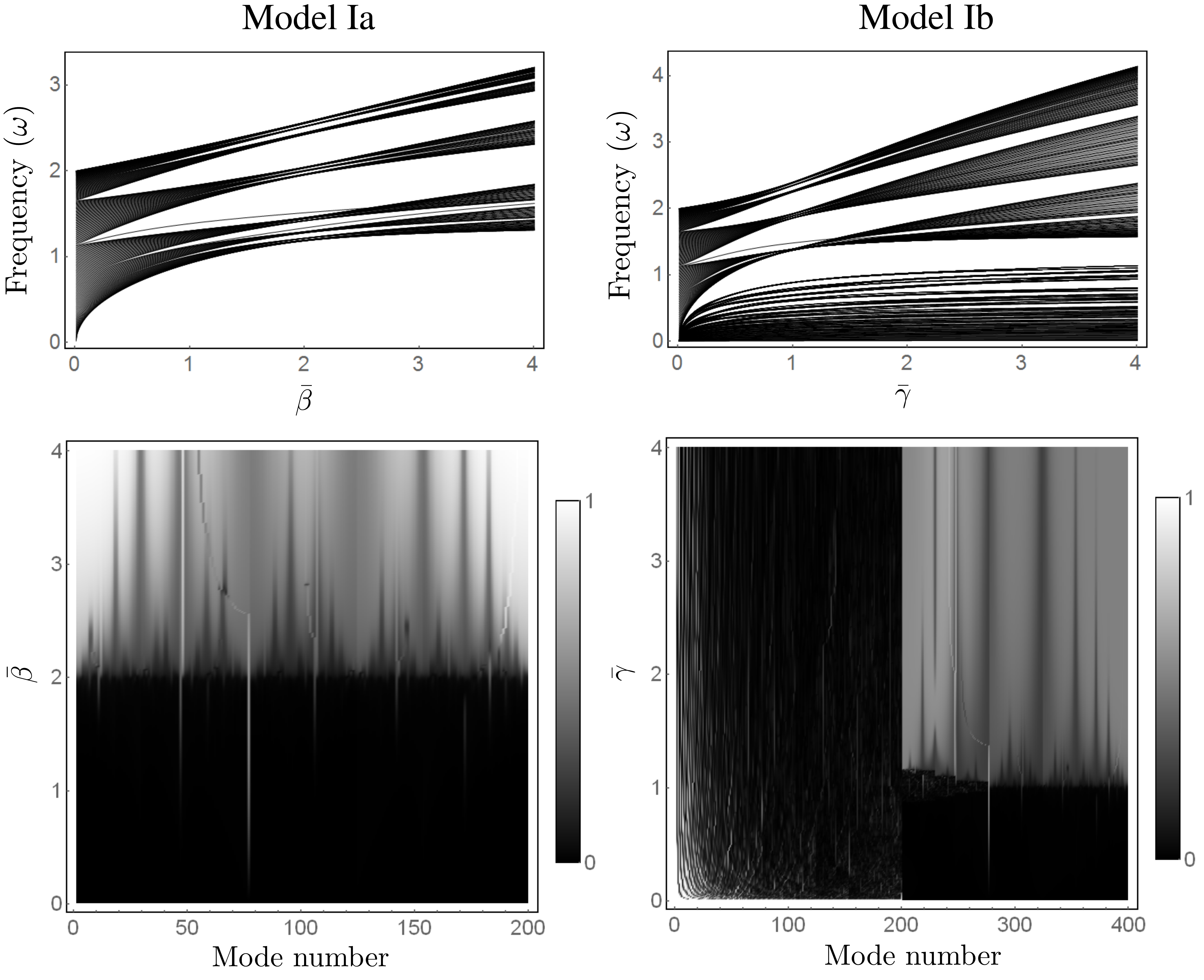}
\caption{(Top) Linear spectrum as a function of the quasiperiodicity parameter and (bottom) inverse participation ratio. We show our results for model Ia in the left column and for model Ib in the right column.
}
\label{figAA1}
\end{figure*}


\subsection{Linear spectrum and localization transition}

In addition to the linear spectrum, which we obtain by solving~\eqref{Eigenvalues}, we also compute the inverse participation ratio (IPR)
\begin{equation}
	P^{-1} = \frac{\sum_n \left[h(u_n,\dot{u}_n)^2 + h(v_n,\dot{v}_n)^2\right]}
	{ \left[\sum_n h(u_n,\dot{u}_n) + h(v_n,\dot{v}_n) \right]^2} \in \left[0,1\right]
\end{equation}
as a measure of the amount of localization of the eigenmodes. For modal analysis, we use $h(u_n,\dot{u}_n) = u_n^2$ and $h(v_n,\dot{v}_n) = v_n^2$. A value of $P^{-1}=1$ accounts for modes when only one particle is vibrating. In contrast, a mode is fully extended if $P^{-1}\rightarrow 0$ as $N\rightarrow \infty$. This provides a qualitative understanding of the nature of the linear modes, and a transition in the IPR also gives a way to quantitatively describe the AA transition.

\begin{figure}
\centering
\includegraphics[height=5cm]{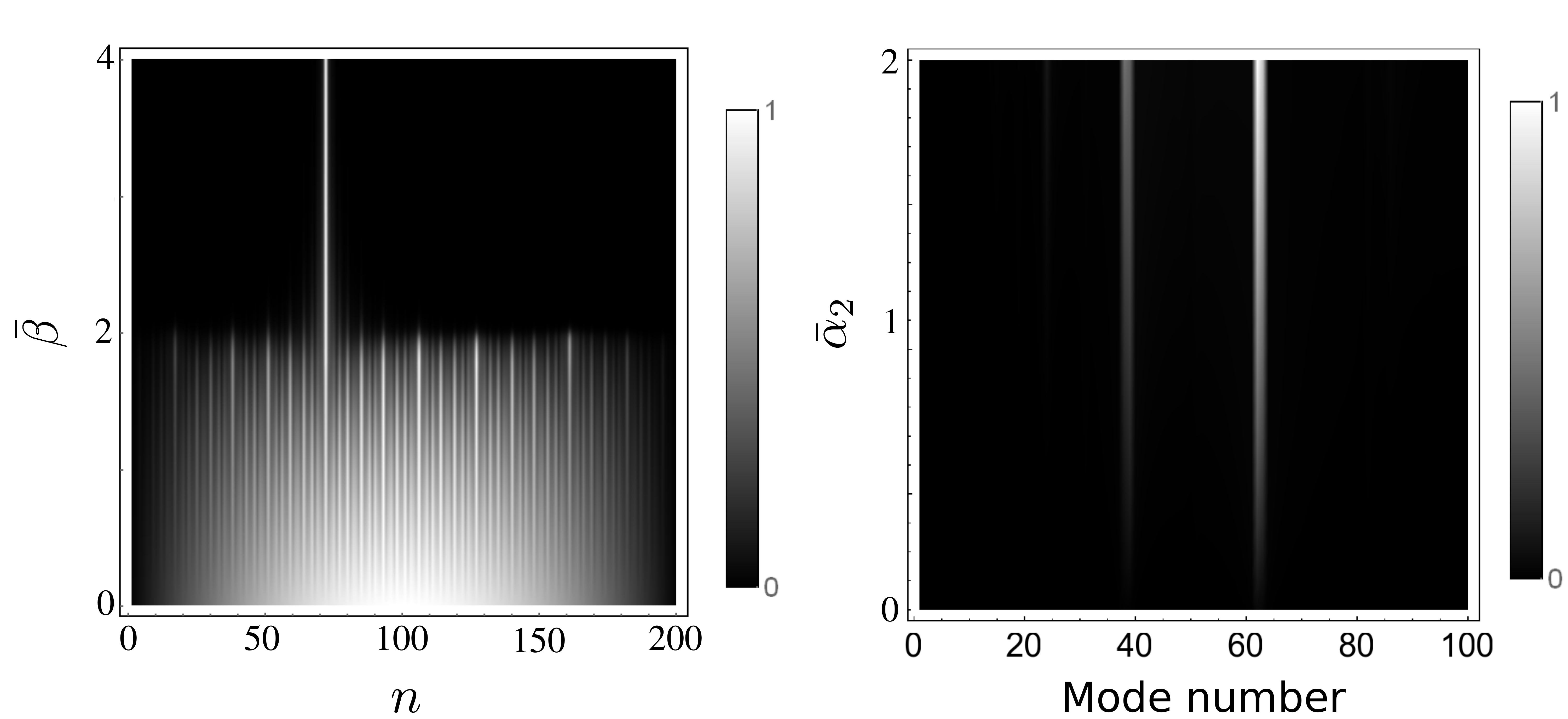}
\caption{(Left) Localization transition as a function of $\bar{\beta}$ for the fundamental mode in model Ia.
(Right) Inverse participation ratio $P^{-1}$ as a function of $\bar{\alpha}_2\in [0,2]$ and $\bar{\alpha}_1=3$ for model II with $N=100$ cylindrical particles.}
\label{figtransition}
\end{figure}

In Fig.~\ref{figAA1}, we show the linear spectrum and IPR as a function of the quasiperiodicity parameter (which is $\bar{\beta}$ in model Ia and $\bar{\gamma}$ in model Ib) for a chain of $N=200$ particles 
for models Ia and Ib. We observe that these two models have a complex 
structure of bands and gaps, with some frequencies that appear
isolated in the gaps and others that form bands that appear to
cluster. Isolated frequencies are associated with modes that are similar to impurity modes. Similar structures of bands and 
gaps have been observed in other physical systems, such
as in optics (see, e.g.,~\cite{OAA,Martinez:PRA2012}). Interestingly, we observe from the IPR that AA transitions occur in granular chains, most prominently in model Ia, where the transition is 
effectively identical to that in the original AA model. This is a consequence of modulating only the on-site potential with an external mechanism, so linearizations of the two systems yield the same equations. If $B_n=1$ and $\gamma_n=0$ for all $n$, the transition occurs at $\bar{\beta}=2$. This differs starkly from the localization properties of the linear modes in the Anderson model, where low-frequency linear modes remain extended independently of the amount of disorder~\cite{DisorderTheo1}. In Fig.~\ref{figtransition}(a), we show the transition to localization in the fundamental mode of model Ia.  
For model Ib, we double the number of modes in the system, because we double the number of DOF by incorporating the internal particles. This system has a very rich spectrum, where the upper part 
(half of the modes) has the same structure as in model Ia, but there is also a bottom part (the other half of the modes) associated with modes that do not undergo the localization transition and consequently are extended independently of the modulation.
This is straightforward to explain by writing the system \eqref{Linear} in terms of in-phase ($x_n = u_n + v_n$) and out-of-phase ($y_n = u_n - v_n$) variables. This yields 
$\dot{x}_n = f(x_n,x_{n\pm 1},y_n,y_{n\pm 1})$ and $\dot{y}_n = g(x_n,x_{n\pm 1},y_n,y_{n\pm 1}) - 2\gamma_n y_n$. 
Only the equations for $\dot{y}_n$ are affected explicitly by the quasiperiodicity. In this case, $2\gamma_n$ enters as a prefactor of $y_n$, instead of $\gamma_n$ as in model Ia. This explains why the localization transition occurs at $\bar{\gamma} = 1$ instead of at $\bar{\gamma} = 2$. Note that modes in the upper part of the spectrum also correspond to out-of-phase modes (between $u_n$ and $v_n$), whereas the bottom part of the spectrum is associated with in-phase modes. The latter
  do not see the quasiperiodicity in practice (because they effectively
  satisfy the original Hertzian dynamics without the MiM contribution),
  so they are generically extended.

In contrast to models Ia and Ib, model II does not have a localization transition; instead, we observe that all modes are extended, except for the ones that are associated with isolated frequencies in the gaps. In Fig.~\ref{figtransition}(b), we show the IPR as a function of $\bar{\alpha}_2$ for model II with $\bar{\alpha}_1=3$. This suggests that, without an on-site potential, one cannot observe this sort of transition in granular chains of cylindrical particles. 
In the future, it will be particularly worthwhile to explore the generality of
this conclusion. Specifically, a relevant question
is whether it is generically the case that it is impossible
for inter-site interactions, modulated by one or more frequencies,
to induce a localization transition in a granular chain.

\section{Hofstadter butterfly}  \label{sec5}

\begin{figure*}
\centering
\includegraphics[height=9cm]{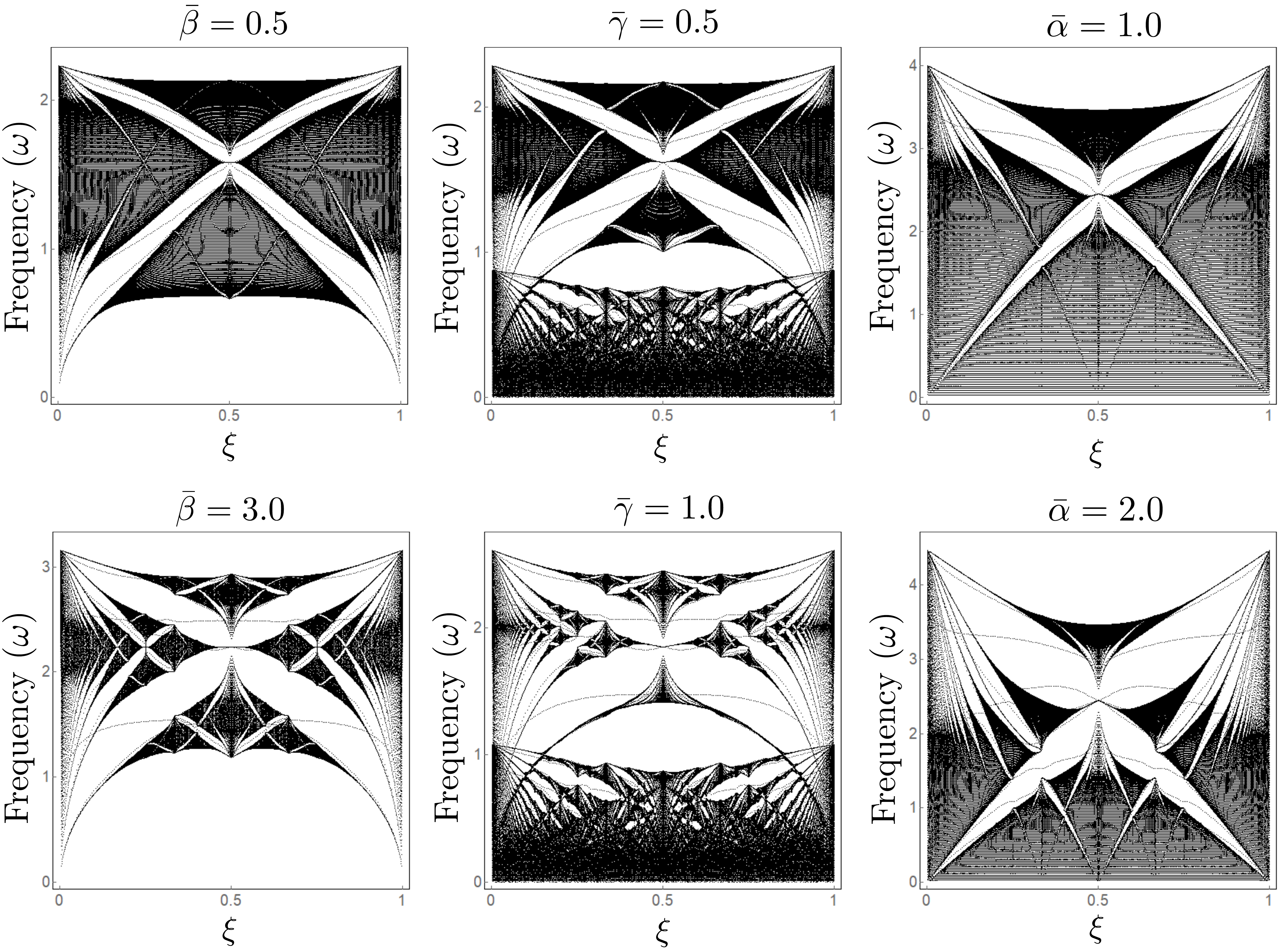}
\caption{Linear spectrum, in the form of a Hofstadter butterfly, as a function of $\xi$ for (left) model Ia, (center) model Ib, and (right) model II.}
\label{AAbutters}
\end{figure*}

Another property of the AA model's spectrum is its fractal nature. To explore this, we compute the spectrum as a function of $\xi$ (see Fig.~\ref{AAbutters}), and we observe a structure that is known as a \textit{Hofstadter butterfly}. The butterfly is a footprint of the spectrum's fractality, and one can see its statistical self-similarity in the figure.  

The Hofstadter butterfly was first predicted in 1976~\cite{butterorigin} in a completely different system: Bloch electrons on 2D lattices and in the presence of an orthogonal magnetic fields. In typical crystals, one needs to use magnetic fields that are at least of the order of several thousands of tesla to observe a Hofstadter butterfly. As a result, it took until 1997 ---  in a microwave system~\cite{buttermicro} --- that a Hofstadter butterfly was observed experimentally. A Hofstadter butterfly was then observed in graphene in 2013~\cite{buttergraphene} and using interacting photons in superconducting qubits in 2017~\cite{sciencebutter}. The possibility to also observe Hofstadter butterflies in granular chains is very exciting, given the simplicity and controllability
of the latter system.

\begin{figure*}
\centering
\includegraphics[height=6.5cm]{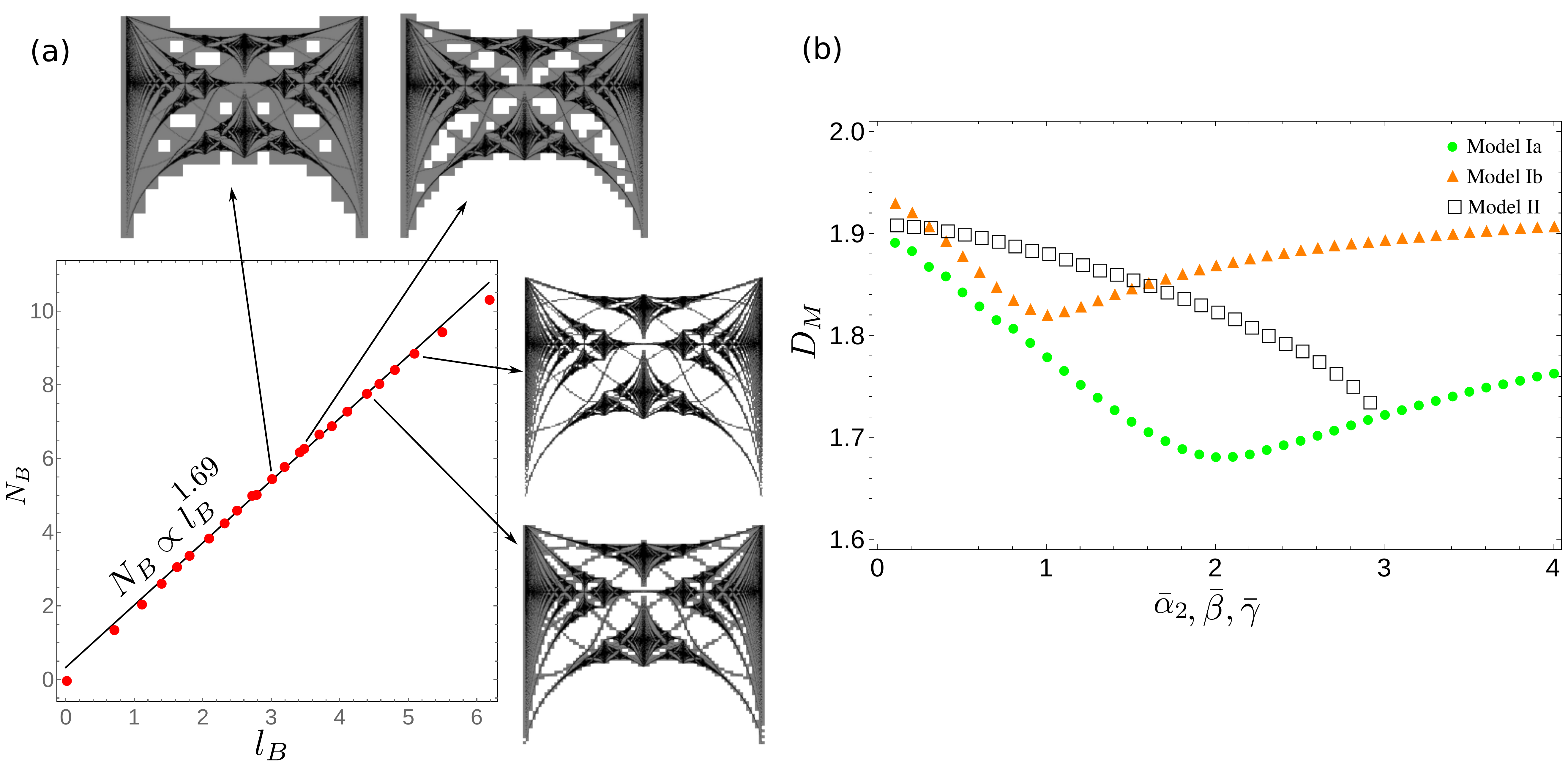}
\caption{(a) Example of the box-counting process that we use to compute the fractal dimension for model Ia and $\bar{\beta}=2$. (b) Minkowski--Bouligand fractal dimension $D_M$ as a function of the 
quasiperiodic parameter $\bar{\beta}$, $\bar{\gamma}$, and $\bar{\alpha}_2$ for models Ia, Ib, and II, respectively. We use the parameter value $\bar{\alpha}_1 = 3$ for model II.}
\label{Boxes}
\end{figure*}

To characterize the self-similarity of the spectrum in the different cases, we compute the Minkowski--Bouligand fractal dimension ($D_m$), which (by Moran's theorem) is the same as the Hausdorff dimension ($D_h$)
for strictly self-similar fractals~\cite{Fractal1,Fractal2}. The procedure to numerically compute $D_M$ is known as \textit{box-counting}, and we proceed as follows. First, we map the Hofstadter 
spectrum into a square of $480\times480$ pixels, we then partition the square
into boxes of characteristic size (side length) $l_B$, and finally we count the number $N_B$ of boxes that include at least one point of the spectrum. We do this procedure for different values of $l_B$; if $\ln(N_B)$ scales 
linearly with $\ln(l_B)$, then the fractal dimension $D_M$ satisfies the relation $N_B \propto l_B^{D_M}$. In practice, one computes $D_M$ as the best fit to $N_B \propto l_B^{D_M}$. For this, we compute a linear regression of the logarithm of the data using gradient descent. In Fig.~\ref{Boxes}(a), we show an example of the box counting for the 
Hofstadter spectrum of model Ia at the localization transition (i.e., when $\bar{\beta}=2$). To show how the band gaps are filled with boxes, we have superimposed the $N_B$ boxes over 
the spectrum for different values of $l_B$. 

We now compute the fractal dimension $D_M$ as a function of the quasiperiodic parameters $\bar{\alpha}_2$, $\bar{\beta}$, and $\bar{\gamma}$ for our three models. We expect $D_M$ to be between $1$ and $2$, because $D_M = 1$ for a line and $D_M = 2$ for a plane. In Fig.~\ref{Boxes}(b), we show the results of our computations.
 We observe that the minimum fractal dimension occurs at the same point as the localization transition for models Ia (at $\bar{\beta}=2$) and Ib (at $\bar{\gamma}=1$). We calculate that $D_M \gtrapprox 1.69$ for model Ia and $D_M \gtrapprox 1.82$ for model Ib. It is interesting to note the similar non-monotonic dependence of the fractal dimension on the model parameters in models Ia and Ib. Presumably, this arises from the aforementioned similarity of the former model and the out-of-phase excitations
 of the latter model. In contrast, in model II, for $\bar{\alpha}_1=3$ and $\bar{\alpha}_2\in(0,3)$, we observe that the fractal dimension decreases monotonically as $\bar{\alpha}_2$ increases.

\section{Energy transport and localization}  \label{sec6}

Another important issue, which one can examine in several ways, is how energy transport is affected by the quasiperiodicity~\cite{transport1,transport2}. For instance, one can compute a 
second moment $m_2$ of the energy distribution
as a function of time to quantitatively characterize the temporal evolution of the energy distribution's 
width~\cite{DisorderTheo1,DisorderTheo2,DisorderExp,m21,m22,m23,m24,m25,m26}. Following recent studies in disordered granular chains~\cite{DisorderExp}, we study the evolution of the energy 
distribution $\{H_n\}_{n=1}^N$ immediately after the impact of a striker against the first particle ($n=1$). 
We write the particles' energy as
\begin{align}
	 H_n &= \frac{\dot{u}_n^2}{2} + \frac{\dot{v}_n^2}{2} + \frac{1}{2}\left\{\frac{2\alpha_n}{5}\left[\delta_n+u_{n-1}-u_{n}\right]_+^{5/2}+ \frac{2\alpha_{n+1}}{5}\left[\delta_{n+1}+u_{n}-u_{n+1}\right]_+^{5/2}\right\}\\
 &\qquad +\frac{\beta_n}{2}u_n^2+\frac{\gamma_n}{2}\left(u_n-v_n\right)^2\,.\nonumber
\end{align}
The total energy is $H = \sum_n H_n$, which is a conserved quantity. Note that, in the context of experiments, one
should also consider dissipation which is neglected here;
see a relevant summary of models thereof
in~\cite{Hertz4}. Importantly, we expect that our results will be robust enough to be observable experimentally even in the presence of a small amount of dissipation. This claim is supported by recent experimental results on the observation of other linear phenomena (e.g., Anderson-like localization~\cite{DisorderExp} and an analog of a Ramsauer--Townsend resonance~\cite{ajm2016b}) in granular chains.

We then compute the second moment   
\begin{equation}
	 m_2(t) = \frac{\sum_n (n-1)^2 H_n}{\sum_n H_n}\,,
\end{equation}
and we estimate a scaling relationship between $m_2$ and $t$. When the scaling is approximated reasonably as a power law --- for which, in exact form, $m_2(t)\sim t^{\eta}$ as $t\rightarrow \infty$ for some
exponent $\eta$ --- one can categorize as \textit{ballistic}
the case when $\eta = 2$, 
\textit{superdiffusive} the one when $\eta \in (1,2)$, \textit{diffusive} when $\eta = 1$, and \textit{subdiffusive} when $\eta \in (0,1)$. If $\eta=0$, we say that there is no diffusion; in other words, all energy remains localized. 
In a perfectly homogeneous granular chain, it is known that energy transport is ballistic. However, when disorder is added to the chain, the dynamics can change drastically, and one can observe different energy transport regimes~\cite{DisorderTheo1,DisorderTheo2,DisorderExp}. These previous studies have focused on the interplay between disorder and nonlinearity. Here, in contrast, we show that even in a strongly precompressed (i.e., almost linear) chain, one can obtain any desired exponent diffusion $\eta \in \left[0,2\right]$ for the energy transport. However, we find that on-site potentials are essential to have localization. 

One advantage of working with quasiperiodic chains instead of disordered ones is that we do not need to compute averages over a large number of realizations to obtain robust insights. Our quasiperiodic chains
are produced in a deterministic way, so given a parameter $\xi$, one gets one specific chain. This enables us to cover the whole parameter space with considerably fewer computations than when studying disordered chains. 
We are also interested in characterizing energy transport in realistic frameworks, so we set $N=21$, which gives a long enough chain to qualitatively capture the nature of transport, at least in several recent experimental and theoretical explorations~\cite{DisorderTheo1,DisorderExp}.

\begin{figure*}
\centering
\includegraphics[height=4.2cm]{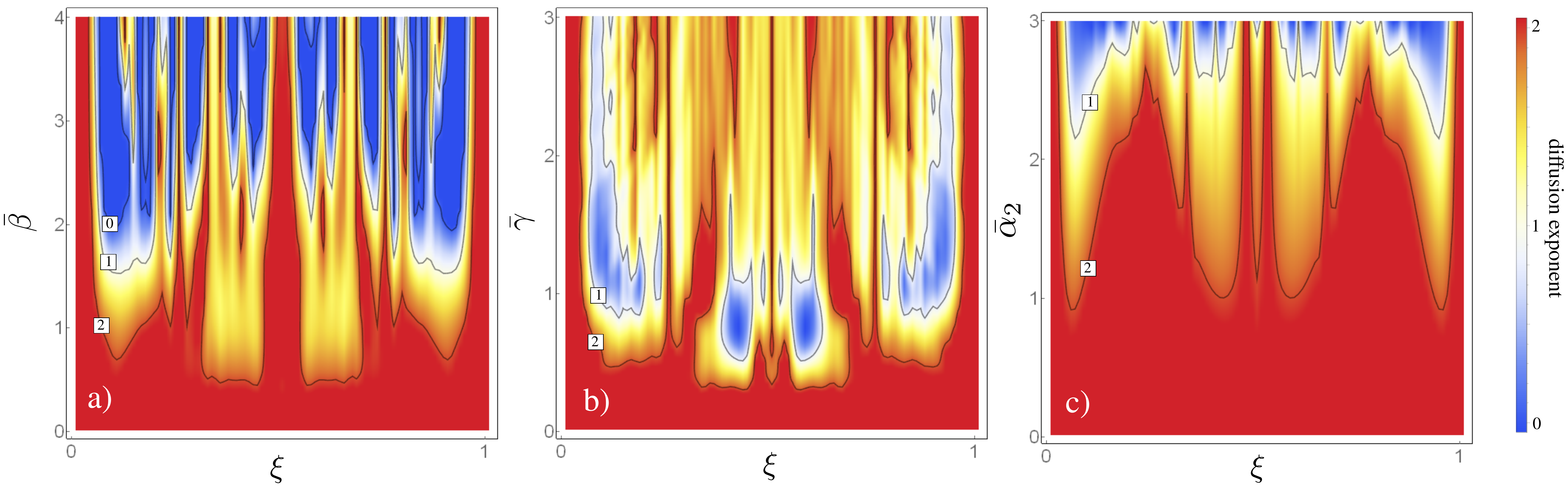}
\caption{Diffusion exponent $\gamma$ for (a) model Ia, (b) model Ib, and (c) model II as a function of the quasiperiodic parameters and $\xi$.}
\label{AAm2}
\end{figure*}

To integrate~\eqref{3.1}, we use a fifth-order explicit Runge--Kutta (RK5) method with a step size of $\mathrm{d}t = 0.01$. We set $\delta_n = 1$ for all $n$. We also set $u_0(t)=u_{N+1}(t)=0$ for all $t$, so we have fixed boundary conditions on both sides. 
We use the stopping criterion for our simulations that either $T = 20$ or that energy reaches the boundary opposite to the one impacted by the striker. We use the former condition to stop the code in cases in which all of the energy is trapped in the form of localized states. In other words, there is no diffusion. This is expected, for instance, 
in model Ia for $\bar{\beta}>2$ (i.e., after the localization transition occurs). However, it is an uncommon scenario 
in model II, which does not have a localization transition. 
In Fig.~\ref{AAm2}, we show our results for energy transport in our three models. 
In models Ia and Ib, we explore the parameters ranges $\bar{\beta}\in[0,4]$ and $\bar{\gamma}\in[0,3]$, respectively, so we can compare energy transport on both sides of the localization transition.
In model II, we consider $\bar{\alpha}_1=3$ and $\bar{\alpha}_2\in[0,3]$. 
For all three models, and for $\xi\in[0,1]$, we can tune the energy transport over from subdiffusive to ballistic behaviors. This allows a great deal of control of the energy-transport properties, and it is remarkable that we are able to do this using a deterministic model. In model Ia, we also observe localization (for which $\eta = 0$) in contrast with observations in disordered granular crystals~\cite{DisorderTheo1,DisorderTheo2,DisorderExp}. This suggests that the inclusion of on-site potentials is crucial for this localization phenomenon.

\section{Conclusions}  \label{sec7}

In this article, we introduced different types of quasiperiodic granular chains that were inspired by the work by Aubry and Andr\'e in condensed-matter physics and by recent developments (cradle and mass-in-mass systems) in the
context of granular lattice systems. We studied the localization and spectral properties of such chains. To achieve each type of quasiperiodic chain, we incorporated spatial modulation (which is incommensurate with the chain's period) into one of the physical parameters.
We proposed three models: models Ia and Ib use spherical particles, in which quasiperiodicity enters via an on-site potential --- either in the form of a local oscillator (as in a cradle) or in the form of a local resonator (for a mass in mass system) --- and model II uses cylindrical particles. In models Ia and Ib, we demonstrated the existence of an analog of the well-known AA transition. However, in model II, we showed that, without an on-site potential and with quasiperiodicity affecting only inter-site
interactions, a localization transition cannot occur. 

We also computed the Hofstadter spectrum for each of the models and studied their fractal properties by computing the Minkowski--Bouligand fractal dimension. In models Ia and Ib, we showed that the minimum fractal dimension $D_M$ of the spectrum coincides with the point at which the localization transition occurs. For model II, we observed that the spectrum's fractal dimension decreases
monotonically as a function of the quasiperiodic parameter. 

Finally, we numerically studied energy transport by exciting the granular chains with a striker. We demonstrated the existence of different regimes --- ranging from ballistic to subdiffusive transport --- as well as 
localization. In contrast to prior work, which achieved such control using a combination of disorder and adjusting the nonlinearity strength~\cite{DisorderTheo1,DisorderTheo2,DisorderExp}, we were able to control the energy transport using a deterministic model in a strongly precompressed (and almost linear) granular chain.

Naturally, it will be particularly valuable to implement some of these ideas in laboratory experiments, as one can then further explore the
role that granular systems can play in the study of quasiperiodic operators~\cite{buttermicro,buttergraphene,sciencebutter}. Achieving an experimental realization of a Hofstadter butterfly in a granular chain would also be very exciting in its own right. Among the settings that we have proposed in this paper, model Ib (i.e., the chain of mass-in-mass particles) is the clearest candidate for
observing a localization transition (in the out-of-phase variables), given that the cradle system has yet to be experimentally realized. Arguably, the woodpile setup of~\cite{jinkyu} may also constitute an excellent playground for such studies. However,
a key consideration, given experimental limitations, is that one seeks to build a chain with as few particles as possible such that one can
(still) observe the relevant phenomenology. 

There are also numerous open issues to explore computationally
and theoretically. Examples include the effect of nonlinearity (e.g., through larger excitation amplitudes) on these modes and their localization, how these phenomena differ for granular crystals in different numbers of dimensions, and others. Relevant extensions will be considered in future studies.


\vskip6pt

\enlargethispage{20pt}


\funding{AJM acknowledges support from CONICYT (BCH72130485/2013).
PGK gratefully acknowledges support from the US-AFOSR under FA9550-17-1-0114.}


\ack{We thank R. Chaunsali, Alain Goriely, Robert MacKay, and Francisco J. Mu\~noz for helpful comments. We also thank the editors of this special issue for their invitation to contribute an article to it.}




\begin{thebibliography}{9}


\bibitem{Janot} Janot C. 1994 \textit{Quasicrystals: A Primer}. 2nd edition, Clarendon Press, Oxford, UK.

\bibitem{NaturalQuasi} Bindi L, Steinhardt PJ, Yao N, Lu PJ. 2009. Natural quasicrystals. \textit{Science} \textbf{324}, 1306.

\bibitem{Definition} Lifshitz R. 2003. Quasicrystals: A matter of definition. \textit{Foundations of Physics} \textbf{33}, 1703.

\bibitem{Penrose} Penrose R. 1978. Pentaplexity. \textit{Eureka} \textbf{39}, 16. 

\bibitem{Fibo} Levine D, Steinhardt PJ. 1984. Quasicrystals: A new class of ordered structures. \textit{Phys. Rev. Lett.} \textbf{53}, 2477.

\bibitem{AAmodel} Aubry S, Andr\'e G. 1980. Analyticity breaking and Anderson localization in incommensurate lattices. \textit{Ann. Isr. Phys. Soc.} \textbf{3}, 133.

\bibitem{Equivalence} Kraus YE, Zilberberg O. 2012. Topological equivalence between the Fibonacci quasicrystal and the Harper model. \textit{Phys. Rev. Lett.} \textbf{109}, 116404.

\bibitem{fishman} Grempel DR, Fishman S, Prange, RE. 1982.
  Localization in an incommensurate potential: An exactly solvable model.
  \textit{Phys. Rev. Lett.} \textbf{49}, 833.

\bibitem{aullbach} Aullbach, C, Wobst A, Ingold, G-L, H{\"a}nggi P,
  Varga I. 2004. Phase-space visualization of a metal-insulator transition,
  \textit{New J. Phys.} \textbf{6}, 70.

\bibitem{flach} Flach S, Ivanchenko M, Khomeriki R.
  2012. Correlated metallic two-particle bound states in quasiperiodic chains.
  \textit{EPL}, \textbf{98}, 66002.

\bibitem{OAA} Lahini Y, Pugatch R, Pozzi F, Sorel M, Morandotti R, Davidson N, Silberberg Y. 2009. Observation of a localization transition in quasiperiodic photonic lattices. \textit{Phys. Rev. Lett.} \textbf{103}, 013901.

\bibitem{refael} Iyer S, Oganesyan V, Refael G, Huse DA.
  2013. Many-body localization in a quasiperiodic system,
  \textit{Phys. Rev. B}, \textbf{87}, 134202.

\bibitem{mastropietro} Mastropietro V.
  2015. Localization of interacting fermions in the Aubry--Andr{\'e} model.
  \textit{Phys. Rev. Lett.}, \textbf{115}, 180401.
  
\bibitem{Hertz1} Nesterenko VF. 2001 \textit{Dynamics of Heterogeneous Materials}. Springer-Verlag, Heidelberg, Germany.

\bibitem{Hertz2} Sen S, Bang J, Avalos E, Doney R. 2008. Solitary waves in the granular chain. \textit{Phys. Rev.} \textbf{462}, 21.

\bibitem{Hertz3} Porter MA, Kevrekidis PG, Daraio C. 2015. Granular crystals: Nonlinear dynamics meets materials engineering. \textit{Phys. Today} \textbf{68}(11), 44.

\bibitem{Hertz4} Chong C, Porter MA, Kevrekidis PG, Daraio C. 2017. Nonlinear coherent structures in granular crystals. \textit{J. Phys.: Condens. Matter} \textbf{29}, 413003.

\bibitem{DisorderTheo1} Mart\'inez AJ, Kevrekidis PG, Porter MA. 2016. Superdiffusive transport and energy localization in disordered granular crystals. \textit{Phys. Rev. E} \textbf{93}, 022902.

\bibitem{DisorderTheo2} Achilleos V, Theocharis G, Skokos Ch. 2016. Energy transport in one-dimensional disordered granular solids. \textit{Phys. Rev. E} \textbf{93}, 022903.

\bibitem{DisorderExp} Kim E, Mart\'inez AJ, Phenisee SE, Kevrekidis PG, Porter MA, Yang J. 2018. Direct measurement of Superdiffusive energy transport in disordered granular chains. \textit{Nat. Commun.} (in press). See \href{https://arxiv.org/abs/1705.08043}{https://arxiv.org/abs/1705.08043}.

\bibitem{ajm2016b} Mart\'inez AJ, Yasuda H, Kim E, Kevrekidis PG,  Porter MA, Yang J. 2016. Scattering of waves by impurities in precompressed granular chains. \textit{Phys. Rev. E} \textbf{93}, 052224.

\bibitem{Nature11} Boechler N, Theocharis G, Daraio C. 2011. Bifurcation based acoustic switching and rectification. \textit{Nat. Mater.} \textbf{10}, 665.

\bibitem{vakbook} Starosvetsky Y, Jayaprakash KR, Arif Hasan M, Vakakis AF.
  2017. Topics on the Nonlinear Dynamics and Acoustics of Ordered Granular
  Media, \textit{World Scientific}, Singapore.

\bibitem{katjarev} Lindenberg K, Harbola U, Romero AH, Lindenberg K.
  2011. Pulse propagation in granular chains. \textit{AIP Conf. Proc.},
  \textbf{1339}, 97

\bibitem{sen2} Przedborski M, Sen S, Harroun TA.
  2017. Fluctuations in Hertz chains at equilibrium.
  \textit{Phys. Rev. E}, \textbf{95}, 032903.

\bibitem{sen3} Han D, Westley M, Sen S.
  2014. Mechanical energy fluctuations in granular chains: The possibility of rogue fluctuations or waves.
  \textit{Phys. Rev. E}, \textbf{90}, 032904.
   
\bibitem{anderson1958} Anderson PW. 1958. Absence of diffusion in certain random lattices. \textit{Phys. Rev.} \textbf{109}, 1492.

\bibitem{Martinez:PRA2012} Mart\'inez AJ, Molina, MI. 2012. Surface solitons in quasiperiodic nonlinear photonic lattices. \textit{Phys. Rev. A} \textbf{85}, 013807.

\bibitem{AABEC} Roati G, D'Errico C, Fallani L, Fattori M, Fort C, Zaccanti M, Modugno G, Modugno M, Inguscio M. 2008. Anderson localization of a non-interacting Bose--Einstein condensate. \textit{Nature} \textbf{453}, 895.

\bibitem{AAPT} Yuce C. 2014. PT symmetric Aubry--Andr{\'e} model. \textit{Phys. Lett. A} \textbf{378}, 2024.

\bibitem{schlag} Goldstein M, Schlag W, Voda M.
  2017. On the spectrum of multi-frequency quasiperiodic Schr\"odinger operators with large coupling,
  arXiv:1708.09711.
  
\bibitem{cradle} James G. 2011. Nonlinear waves in Newton's cradle and the discrete p-Schr\"odinger equation. \textit{Math. Models methods Appl. Sci.} \textbf{21}, 2335.

\bibitem{cuevas} James G, Kevrekidis PG, Cuevas J. 2013.
  Breathers in oscillator chains with Hertzian interactions,
  \textit{Physica D}, \textbf{251}, 39.
  
\bibitem{mim} Huang H, Sun C, Huang G. 2009. On the negative effective mass density in acoustic metamaterials. \textit{Ing. J. Eng. Sci.} \textbf{47}, 4.

\bibitem{mim1} Kevrekidis PG, Vainchtein A, Serra Garcia M, Daraio C.
  2013. Interaction of traveling waves with mass-with-mass defects within a Hertzian chain. \textit{Phys. Rev. E}, \textbf{87}, 042911.

\bibitem{mim2} Bonanomi L, Theocharis G, Daraio C.,
  2015. Wave propagation in granular chains with local resonances.
  \textit{Phys. Rev. E}, \textbf{91}, 033208.
  
\bibitem{cili1} Khatri D, Ngo D, Daraio C. 2012. Highly nonlinear solitary waves in chains of cylindrical particles. \textit{Granular Matter} \textbf{14}, 63.

\bibitem{fli} Li F, Chong C, Yang J, Kevrekidis PG, Daraio C.
  2014. Wave transmission in time- and space-variant helicoidal phononic crystals.
  \textit{Phys. Rev. E}, \textbf{90}, 053201.  
  
\bibitem{Mulansky:NJP2013} Mulansky M, Pikovsky A. 2013. Energy spreading in strongly nonlinear disordered lattices. \textit{New J. Phys.} \textbf{15}, 053015.

\bibitem{haitao} Xu H, Kevrekidis PG, Stefanov A. 2015.
  Traveling waves and their tails in locally resonant granular systems,
  \textit{J. Phys. A}, \textbf{48}, 195204.

\bibitem{physd} Vorotnikov K, Starosvetsky Y, Theocharis G, Kevrekidis PG.
  2018. Wave propagation in a strongly nonlinear locally resonant granular crystal. \textit{Physica D}, \textbf{365}, 27.

\bibitem{anna1} Liu L, James G, Kevrekidis P, Vainchtein A.
  2016. Strongly nonlinear waves in locally resonant granular chains.
  \textit{Nonlinearity}, \textbf{29}, 3496.  

  \bibitem{anna2} Liu L, James G, Kevrekidis P, Vainchtein A.
    2016. Breathers in a locally resonant granular chain with precompression.
\textit{Physica D}, \textbf{331}, 27.    

\bibitem{jinkyu} Kim E, Li F, Chong C, Theocharis G,
  Yang J, Kevrekidis PG.
  2015. Highly nonlinear wave propagation in elastic woodpile periodic structures, \textit{Phys. Rev. Lett.}, \textbf{114}, 118002.
  
\bibitem{DLMF} Digital Library of Mathematical Functions (release 1.0.10). 2015. National Institute of Standards and Technology. Available at {\tt http://dlmf.nist.gov/}.

\bibitem{Johnson:Book} Johnson KL. 1987. {\it Contact Mechanics}. Cambridge University Press, New York, USA.

\bibitem{butterorigin} Hofstadter DR. 1976. Energy levels and wavefunctions of Bloch electrons in rational and irrational magnetic fields. \textit{Phys. Rev. B} \textbf{14}, 2239.

\bibitem{buttermicro} Kuhl U, St\"ockmann H-J. 1998. Microwave realization of the Hofstadter butterfly. \textit{Phys. Rev. Lett.} \textbf{80}, 3232.

\bibitem{buttergraphene} Dean CR, Wang L, Maher P, Forsythe C, Ghahari F, Gao Y, Katoch J, Ishigami M, Moon P, Koshino M, Taniguchi T, Watanabe K, Shepard KL, Hone J, Kim P. 2013. Hofstadter's butterfly and the fractal quantum Hall effect in moir\'e superlattices. \textit{Nature} \textbf{497}, 598.

\bibitem{sciencebutter} Roushan P, Neill C, Tangpanitanon J, Bastidas VM, Megrant A, Barends R, Chen Y, Chen Z, Chiaro B, Dunsworth A, Fowler A, Foxen B, Giustina M, Jeffrey E, Kelly J, Lucero E, Mutus J, Neeley M, Quintana C, Sank D, Vainsencher A, Wenner J, White T, Neven H, Angelakis DG, Martinis J. 2017. Spectroscopic signature of localization with interacting photons in superconducting qubits. \textit{Science} \textbf{358}, 1175.

\bibitem{Fractal1} Falconer K. 1990 \textit{Fractal geometry: mathematical foundations and applications}. John Wiley \& Sons, Chichester, UK.

\bibitem{Fractal2} Schroeder M. 1991 \textit{Fractals, Chaos, Power Laws: Minutes from an Infinite Paradise}. W. H. Freeman and Company, New York, USA.


\bibitem{transport1} Kramer B, MacKinnon A. 1993. Localization: Theory and experiment. \textit{Rep. Prog. Phys.} \textbf{56}, 1469.

\bibitem{transport2} Laptyeva TV, Ivanchenko MV, Flach S. 2014. Nonlinear lattice waves in heterogeneous media. \textit{J. Phys. A} \textbf{47}, 493001.

\bibitem{m21} Datta PK, Kundu K. 1995. Energy transport in one-dimensional harmonic chains. \textit{Phys. Rev. B} \textbf{51}, 6287.

\bibitem{m22} Lepri S, Schilling R, Aubry S. 2010. Asymptotic energy profile of a wave packet in disordered chains. \textit{Phys. Rev. E} \textbf{82}, 056602.

\bibitem{m23} Naether U, Rojas-Rojas S, Mart\'inez AJ, St\"utzer S, T\"unnermann A, Nolte S, Molina MI, Vicencio RA, Szameit A. 2013. Enhanced distribution of a wave-packet in lattices with disorder and nonlinearity. \textit{Opt. Express} \textbf{21}, 927.

\bibitem{m24} Garc\'ia-Mata I, Shepelyansky DL. 2009. Delocalization induced by nonlinearity in systems with disorder. \textit{Phys. Rev. E} \textbf{79} 026205.

\bibitem{m25} Laptyeva TV, Bodyfelt JD, Krimer DO, Skokos Ch, Flach S. 2010. The crossover from strong to weak chaos for nonlinear waves in disordered systems. \textit{Europhys. Lett.} \textbf{91}, 30001.

\bibitem{m26} Rojas-Rojas S, Morales-Inostroza L, Naether U, Xavier GB, Nolte S, Szameit RA, Vicencio R, Lima G, Delgado A. 2014. Analytical model for polarization-dependent light propagation in waveguide arrays and applications. \textit{Phys. Rev. A} \textbf{90}, 063823.

\end{thebibliography}
\end{document}